\documentclass[preprint,aps]{revtex4}
\usepackage{psfig}
\begin{document}
\title{Towards a Non-extensive Random Matrix Theory.}
\author{John Evans}
\email{evans@physics.ucf.edu}
\author{Fredrick Michael}
\email{fnm@physics.ucf.edu}
\affiliation{Department of Physics, University of Central Florida, Orlando, FL
32816-2385}
\date{July 16, 2002}
\begin{abstract}
In this article the statistical properties of symmetrical
random matrices whose elements are drawn from a $q$-parametrized
non-extensive statistics power-law distribution are investigated. In the limit as $%
q\rightarrow 1$ the well known Gaussian orthogonal ensemble (GOE) results
are recovered. The relevant level spacing distribution is derived and one
obtains a suitably generalized non-extensive Wigner distribution which depends
on the value of the tunable non-extensivity parameter $q$. This non-extensive Wigner distribution 
can be seen to be a one-parameter level-spacing distribution that allows one to interpolate 
between chaotic and nearly integrable regimes.
\end{abstract} 
\maketitle

\bigskip
The Hamiltonian $M\times M$ matrix that is examined is symmetric and real. In order
to simplify the derivation $M$ is restricted to $M=2$. The derivation of the $2\times 2$
random matrix statistics follows the pedagogical approach of Brody {\itshape{et al}} \cite{brody1,mehta1}. 
A discussion of random matrix theory as applied to quantum chaos and integrability is found in 
\cite{haake1,stockmann1}.
The Hamiltonian matrix is 
\begin{center}
\begin{equation}
H=\left[ 
\begin{array}{cc}
H_{11} & H_{12} \\ 
H_{12} & H_{22}
\end{array}
\right] .  \label{ham1}
\end{equation}
\end{center}
The first task in random matrix theory is to obtain the probability of
occurrence of the individual elements in the ensemble. This is obtained in 
this derivation from the maximum entropy approach \cite{tsallis1}. 
Also, in order to guide the subsequent non-extensive statistics derivation, 
the Gaussian distributed random matrix results are derived first and then 
generalized to the non-extensive case.

The question of the probability of occurrence of the individual elements of
the Hamiltonian matrix in the maximum entropy approach is one of obtaining
the central moments, i.e. the means, variances of the matrix elements. In order 
to obtain the most likely (least biased)
probability density for the occurrence of the matrix elements the entropy of the 
system is 
maximized $\left\langle S\right\rangle $ given the set of 
$N$ constraints (the observables) $\sum\limits_{i=1}^{N}\left\langle
\vartheta _{i}\right\rangle $ 
\begin{center}
\begin{eqnarray}
\left\langle S\right\rangle  &=&-\int P(H_{11},H_{22},H_{12})\text{ }%
lnP(H_{11},H_{22},H_{12})dH_{11}dH_{22}dH_{12},  \nonumber \\
\left\langle \left( H_{11}^{2}+H_{22}^{2}+2H_{12}^{2}\right) \right\rangle 
&=&\int \left( H_{11}^{2}+H_{22}^{2}+2H_{12}^{2}\right) \text{ }%
P(H_{11},H_{22},H_{12})dH_{11}dH_{22}dH_{12}  \nonumber \\
&=&\sigma ^{2}.
\end{eqnarray}
\end{center}

The maximization of the entropy is 
\begin{equation}
\delta \left\langle S\right\rangle -\delta \lbrack \beta \left\langle \left(
H_{11}^{2}+H_{22}^{2}+2H_{12}^{2}\right) \right\rangle ]\equiv 0,
\end{equation}
and here $\beta $ is a Lagrange multiplier which is related to the variance
by $\beta =\frac{1}{2\sigma ^{2}}$. The maximization yields the least biased
probability density ($A$ is the normalization)
\begin{eqnarray}
P(H) &=&P(H_{11},H_{22},H_{12})  \nonumber \\
&=&A\text{ }e^{-\beta \left( H_{11}^{2}+H_{22}^{2}+2H_{12}^{2}\right) }.
\end{eqnarray}
The probability density is seen to be a Gaussian and satisfies the
normalization condition 
\begin{equation}
\int P(H)dH_{11}dH_{22}dH_{12}=1.
\end{equation}
Furthermore, this form of distribution has the
following properties that are of importance in the random matrix theory. 

(i) The probability density of the statistically independent random matrix elements 
is factorizable. This is written as
\begin{equation}
P(H_{11},H_{22},H_{12})=P(H_{11})P(H_{22})P(H_{12}),
\end{equation}
and can be seen by inspection to be true. This property also indicates that
there are no correlations in the probabilty density as the individual elements 
are perforce statistically independent. 

(ii) There is a related issue of
extensivity (additivity). The entropy measure for the system is formed from
the addition of the entropies of the parts such that (omitting constants) 
\begin{eqnarray}
S &=&\text{ }lnP(H_{11},H_{22},H_{12})=ln[P(H_{11})P(H_{22})P(H_{12})] 
\nonumber \\
&=&lnP(H_{11})+lnP(H_{22})+lnP(H_{12}).
\end{eqnarray}

(iii) The probability density is invariant under orthogonal transformations of
the form 
\begin{eqnarray}
H^{\prime } &=&O^{T}HO,  \nonumber \\
O &=&\left[ 
\begin{array}{cc}
cos(\theta ) & sin(\theta ) \\ 
-sin(\theta ) & cos(\theta )
\end{array}
\right] .
\end{eqnarray}
This transformation yields the following transformed variables in the
infinitesimal limit of $\theta \rightarrow 0$%
\begin{eqnarray}
H_{11}^{\prime } &=&H_{11}-2\theta H_{12}  \nonumber \\
H_{22}^{\prime } &=&H_{22}+2\theta H_{12}  \nonumber \\
H_{12}^{\prime } &=&H_{12}+\theta (H_{11}-H_{22}).
\end{eqnarray}
A transformation of the probability density is written as 
\begin{equation}
P(H)dH=P(H^{\prime })dH^{\prime },
\end{equation}
and a calculation of the determinant of the Jacobian of the orthogonal
transformation  $\frac{dH^{\prime }}{dH}=det(J)=1$ yields
\begin{equation}
P(H)=P(H^{\prime }).
\end{equation}
This probability density is invariant under orthogonal
transformations, and the level statistics obtained is  known as The Gaussian orthogonal
ensemble (GOE). Unitary and simplectic
transformations and symmetry considerations yield the Gaussian unitary ensemble (GUE) and
the Gaussian simplectic ensemble (GSE) in a related fashion, however these cases of 
symmetries and degeneracies will not be discussed further in
this article.

\bigskip 

The extensive Gaussian random matrix theory can be generalized by examining
the non-extensive, or $q$-parametrized
entropy \cite{tsallis1,wang1} (equivalently, an incomplete information measure). For systems with 
statistically dependent (say, Hamiltonian matrix elements) variables the
joint probability decomposition is 
\begin{equation}
P(H_{i},H_{j})=P(H_{i}\mid H_{j})P(H_{j}),
\end{equation}
which gives the pseudo-additive entropy 
\begin{eqnarray}
S_{q}(H_{i},H_{j}) &=&S_{q}(H_{j})+S_{q}(H_{i}\mid
H_{j})+(1-q)S_{q}(H_{j})S_{q}(H_{i}\mid H_{j}),  \nonumber \\
S_{q} &=&-ln_{q}P=-\frac{P^{1-q}-1}{1-q}.
\end{eqnarray}
It is known \cite{rajagopal1} that the Tsallis entropy satisfies this
condition, and the resulting probability will be of the power-law form.
The $q$-logarithm is $ln_{q}X=-\frac{1-X^{1-q}}{(q-1)}$. In
the limit as $q\rightarrow 1$ the usual form of the natural
logarithm and thus the extensive statistics and its exponential (Gaussian)
distributions is recovered.

The entropy to be maximized given the constraints is then  
\begin{eqnarray}
\left\langle S\right\rangle _{q} &=&-\frac{1-\int
P^{q}(H_{11},H_{22},H_{12})dH_{11}dH_{22}dH_{12}}{(q-1)}\text{ },  \nonumber \\
\left\langle \left( H_{11}^{2}+H_{22}^{2}+2H_{12}^{2}\right) \right\rangle
_{q} &=&\int \left( H_{11}^{2}+H_{22}^{2}+2H_{12}^{2}\right) \text{ }%
P^{q}(H_{11},H_{22},H_{12})dH_{11}dH_{22}dH_{12}  \nonumber \\
&=&\sigma _{q}^{2},
\end{eqnarray}
and which is subject to the extra normalization condition 
\begin{center}
\begin{equation}
P(H_{11},H_{22},H_{12})dH_{11}dH_{22}dH_{12}=1.
\end{equation}
\end{center}
 
The maximization is then
similar to the Gaussian case, and the variation given the
constraints is 
\begin{equation}
\delta \left\langle S\right\rangle _{q}-\delta \lbrack \beta \left\langle
\left( H_{11}^{2}+H_{22}^{2}+2H_{12}^{2}\right) \right\rangle _{q}]\equiv 0,
\end{equation}
which upon solving yields the least biased probability density
\begin{equation}
P(H_{11},H_{22},H_{12})=A\text{ }\left( 1+\beta
(q-1)(H_{11}^{2}+H_{22}^{2}+2H_{12}^{2})\right) ^{\frac{-1}{q-1}}.
\end{equation}
This is then the Tsallis power-law form of the probability density for the $%
2\times 2$ random matrix elements. Having obtained the statistics of the
random matrix, the explicit form of the non-extensive
Wigner distribution is derived. This non-extensive Wigner distribution will
be the level spacing statistics for a system with statistically dependent random 
matrix elements.

The eigenvalues of the $2\times 2$ Hamiltonian matrix, Eq.(\ref{ham1}), are
given by
\begin{equation}
E_{\pm }=\frac{1}{2}\left( H_{11}+H_{22}\right) \pm \frac{1}{2}\left[ \left(
H_{11}-H_{22}\right) +4H_{12}^{2}\right] ^{\frac{1}{2}},
\end{equation}
which can be written in terms of a diagonal matrix
\begin{equation}
D=\left[ 
\begin{array}{cc}
E_{+} & 0 \\ 
0 & E_{-}
\end{array}
\right] .
\end{equation}
This matrix is related to the Hamiltonian matrix by another orthogonal
transformation $H=\Omega D\Omega ^{T}$ where $\Omega $ is
\begin{equation}
\Omega =\left[ 
\begin{array}{cc}
cos(\phi ) & -sin(\phi ) \\ 
sin(\phi ) & cos(\phi )
\end{array}
\right] .
\end{equation}
This transformation then relates the variables as
\begin{eqnarray}
H_{11} &=&E_{+}cos^{2}(\phi )+E_{-}sin^{2}(\phi )  \nonumber \\
H_{22} &=&E_{+}sin^{2}(\phi )+E_{-}cos^{2}(\phi )  \nonumber \\
H_{12} &=&(E_{+}-E_{-})cos(\phi )sin(\phi ).
\end{eqnarray}
Next the probability densities are transformed directly as
\begin{equation}
P(H_{11},H_{22},H_{12})dH_{11}dH_{22}dH_{12}=P(E_{+},E_{-},\phi
)dE_{+}dE_{-}d\phi ,
\end{equation}
and calculating the determinant of the Jacobian of the transformation gives
\begin{equation}
det(J)=det\frac{\partial (H_{11},H_{22},H_{12})}{\partial (E_{+},E_{-},\phi )%
}=E_{+}-E_{-}\text{ }.
\end{equation}
Rewriting the argument of the probability density in terms of the new
variables 
\begin{equation}
H_{11}^{2}+H_{22}^{2}+2H_{12}^{2}=E_{+}^{2}-E_{-}^{2},
\end{equation}
shows that the probability density is clearly independent of $\phi $ . The transformed probability can
then be written as ($P(E_{+},E_{-},\phi )=P(H)det(J)$) 
\begin{equation}
P(E_{+},E_{-})=A\text{ }\left( E_{+}-E_{-}\right) \text{  }\left( 1+\beta
(q-1)(E_{+}^{2}+E_{-}^{2})\right) ^{\frac{-1}{q-1}}.
\end{equation}

In order to obtain the non-extensive Wigner distribution, the
variables are recast in terms of the level spacing and an auxiliary variable that plays
the role of a `center of mass' energy coordinate  
\begin{eqnarray}
s &=&E_{+}-E_{-}  \nonumber \\
z &=&\frac{E_{+}+E_{-}}{2}.
\end{eqnarray}
Substitution of these variables into the probability density results in  
\begin{equation}
P(s,z)=A\text{ }s\text{ }\left( 1+\beta (q-1)(\frac{s^{2}}{2}+2z^{2})\right)
^{\frac{-1}{q-1}},
\end{equation}
and integration over the auxiliary variable $z$ yields the level spacing
distribution 
\begin{equation}
P(s)=A\text{ }\frac{\sqrt{\frac{\pi }{2}}\Gamma \left[ \frac{1}{q-1}-\frac{1%
}{2}\right] }{2\sqrt{\beta (q-1)}\Gamma \left[ \frac{1}{q-1}\right] }\text{ }%
s\text{ }\left( 1+\beta (q-1)(\frac{s^{2}}{2})\right) ^{\frac{-1}{q-1}+\frac{%
1}{2}}.
\end{equation}
Next the level spacing distribution $P(s)$ is normalized to a mean level spacing of one 
$\left\langle s\right\rangle = 1$ and one obtains the normalized Wigner distribution
\begin{center} 
\begin{equation}
P\left( s\right) _{W}=\beta \left( q-1\right) \frac{\Gamma \left( \frac{1}{%
q-1}-\frac{1}{2}\right) }{\Gamma \left( \frac{1}{q-1}-\frac{3}{2}\right) }%
\text{ }s\text{ }\left( 1+\beta \left( q-1\right) s^{2}/2\right) ^{\frac{-1}{q-1}+\frac{1}{2}%
},
\end{equation}
\end{center}
where $\beta$ is given by 
$\beta \left( q\right) =\frac{1}{2}\frac{1}{q-1}\left( \frac{\Gamma \left( 
\frac{1}{q-1}-2\right) }{\Gamma \left( \frac{1}{q-1}-\frac{3}{2}\right) }%
\right) ^{2}$.

The behavior of the non-extensive Wigner distribution is obtained by plotting
its values for a range of the level spacing $s$, given the `inverse
variance' $\beta $ and the nonextensivity parameter $q$. In Fig.1. the $q$ dependence 
of $\beta $ is plotted for values of $q$ between $1<q<1.4 $. In Fig.2. the Poisson (solid), 
extensive (long-short dashing) and 
non-extensive Wigner (short dashing)
distributions are plotted for a low non-extensivity parameter $q$ value of $q=1.01$. 
The non-extensive distribution 
is nearly superimposed on the extensive Wigner distribution as is expected for $q->1$. In 
Fig.3. the Poisson (solid), 
extensive (long-short dashing) and 
non-extensive Wigner (short dashing) are plotted for a high value of the non-extensivity 
parameter $q=1.38$. Here the distribution is greatly shifted and approaches the Poisson level 
statistics distribution.  

\begin{center}
\begin{figure}
\psfig{file=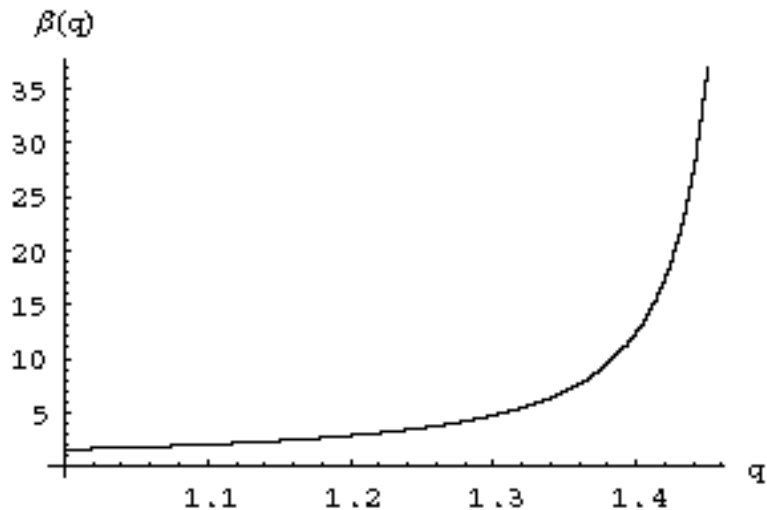}
\caption{$\beta$ Vs. $q$}\label{betafig}
\end{figure}
\end{center}

\begin{figure}
\begin{center}
\psfig{file=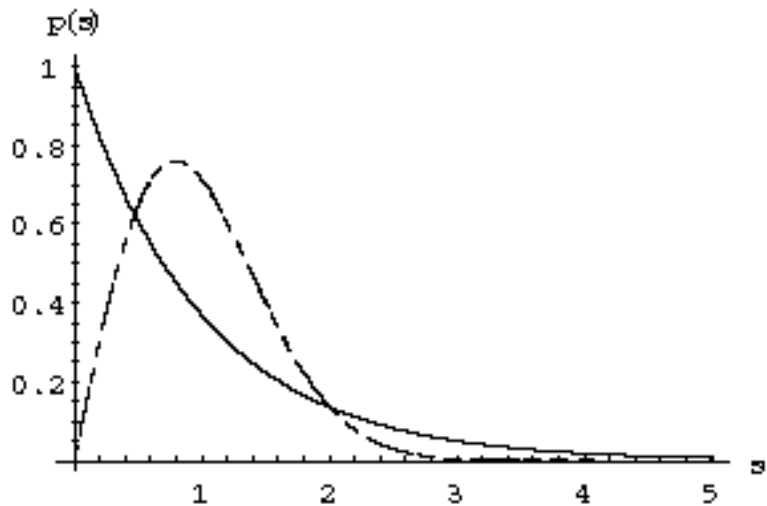}
\caption{$P\left( s\right) _{W}$ Vs. $s$, $q=1.01$. The non-extensive and extensive Wigner 
distributions are nearly super-imposed. The Poisson distribution is plotted using a solid line.}\label{lowq}
\end{center}
\end{figure}

\begin{figure}
\begin{center}
\psfig{file=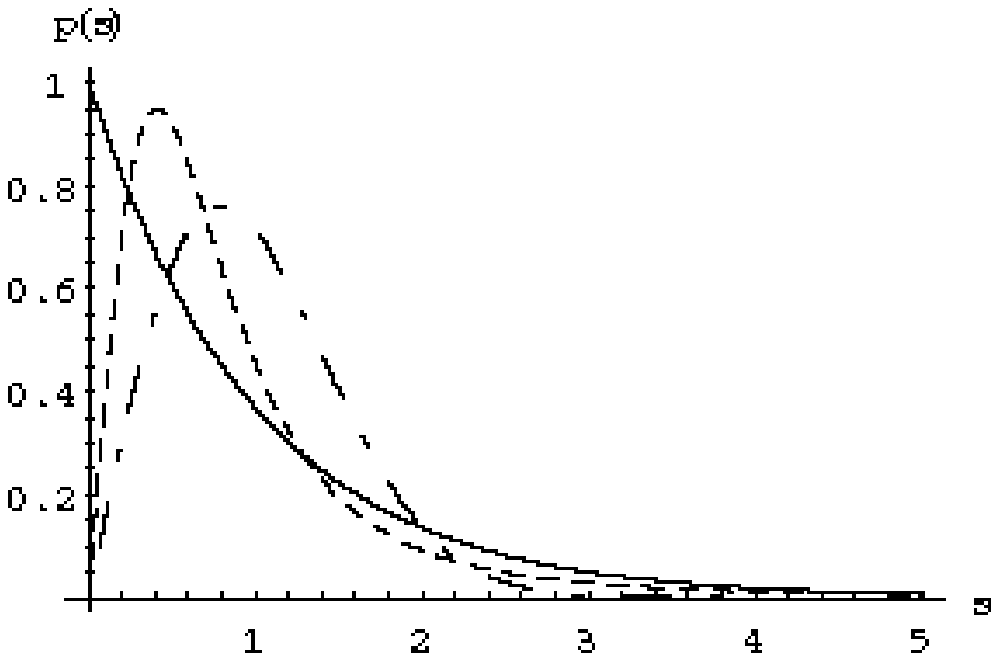}
\caption{$P\left( s\right) _{W}$ Vs. $s$, $q=1.38$. The extensive distribution has been plotted 
with long-short dashing. The non-extensive Wigner distribution is plotted with the short dashes. The 
Poisson distribution is plotted using a solid line.}\label{highq}
\end{center}
\end{figure} 

In this letter the Gaussian orthogonal ensemble (GOE) results for a 
$2\times 2$ random Hamiltonian matrix is generalized to the case
of the non-extensive statistics and the resultant power-law distributions. 
A derivation of the subsequent
level spacing statistical distribution, the non-extensive Wigner
distribution, is given. This derivation is obtained by maximizing the Tsallis non-extensive 
entropy for the $2\times 2$
symmetrical random matrix elements. This can be straight-forwardly generalized to $%
M\times M$ matrices. The resultant non-extensive Wigner level-spacing distribution is 
$q$-parametrized and allows for a smooth interpolation between the extensive Wigner 
distribution and the regime where the level statistics are given by a Poisson distribution. 
In future work it will be interesting to apply these results to Hamiltonians of mixed 
systems between regular and chaotic regimes where deviations from the Wigner statistics 
become pronounced.  

\bigskip 

One of the authors, F. Michael, wishes to acknowledge support from the NSF
through grant number DMR99-72683.

\end{document}